\documentclass[conference]{IEEEtran}
\IEEEoverridecommandlockouts
\usepackage{cite}
\usepackage[ruled]{algorithm2e} 
\usepackage{qtree}
\usepackage{algpseudocode}
\usepackage{amsmath,amssymb,amsfonts}
\usepackage{enumitem}
\usepackage{subcaption}
\usepackage{booktabs}
\usepackage{graphicx}
\usepackage{textcomp}
\usepackage{xcolor}
\def\BibTeX{{\rm B\kern-.05em{\sc i\kern-.025em b}\kern-.08em
    T\kern-.1667em\lower.7ex\hbox{E}\kern-.125emX}}
\begin{document}

\title{Discovering Interesting Subgraphs in Social Media Networks}

\author{\IEEEauthorblockN{Subhasis Dasgupta}
\IEEEauthorblockA{\textit{San Diego Supercomputer Center} \\
\textit{University of California San Diego}\\
La Jolla, USA \\
sudasgupta@ucsd.edu}
\and
\IEEEauthorblockN{Amarnath Gupta}
\IEEEauthorblockA{\textit{San Diego Supercomputer Center} \\
\textit{University of California San Diego}\\
La Jolla, USA \\
a1gupta@ucsd.edu}
}
\maketitle

\begin{abstract}
Social media data are often modeled as heterogeneous graphs with multiple types of nodes and edges. We present a discovery algorithm that first chooses a ``background'' graph based on a user's analytical interest and then automatically discovers subgraphs that are structurally and content-wise distinctly different from the background graph. The technique combines the notion of a \texttt{group-by} operation on a graph and the notion of subjective interestingness, resulting in an automated discovery of interesting subgraphs. Our experiments on a socio-political database show the effectiveness of our technique.
\end{abstract}


\section{Introduction}
\label{sec:intro}
Social Media refers to a set of web-based applications where users create their own profiles and identities, posting their own data content, perform online interactions through operations like ``following", ''re-posting'', ``commenting'' on each other's content, forming interest-based subgroups, and in the process, create a social network amongst users and themes of interest. Social media is often modeled as evolving graphs -- graphs where the nodes represent entities (e.g., users. geographic objects), themes (e.g., hashtags), content (e.g., posts, URLs) and so forth, while the edges represent relationships such as ``a post \textit{commenting on}  another'', ``a user having a \textit{friendship with} another'', ``a post \textit{containing} a hashtag'' and so forth. For some applications, computationally derived edges are used -- for example, hashtag co-occurrence (i.e., the fact that a pair of hashtags has appeared in the same post) is a commonly used derived edge. A typical social media graph has both node properties (e.g., date of a post) and edge properties (e.g., co-occurrence count, the time-interval over which a friendship relationship holds). In addition, a social media graph may have named subgraphs such as user-defined sub-communities (e.g., a Facebook group) which may have their own properties (e.g., the ``privacy level'' of the group). 

A formal data model for a social media network can be specified by extending the well-known property graph model where nodes and edges can have types and each node and edge has its own set of properties. In the extended version, proposed in \cite{junghanns2015gradoop}, subgraphs of a data graph can also be named and  modeled as first class data objects. In the current work, we do not develop a new data model, instead, we assume the EPGM model by \cite{junghanns2015gradoop}, and customize it to suit our requirements.

This paper investigates a technique to discover ``interesting subgraphs'' from a \textit{Social Media Graph}. We formalize the notion of ``interestingness'' in Section \ref{sec:formulation}. Informally, a subgraph of a social media network is ``interesting'' if the subgraph has a structure and content that is sufficiently different from the rest of some reference social media network. There are many reasons why a subgraph would be different from the overall tweet graph. Consider the first tweet shown in Table \ref{tab:tweets} -- the entire tweet has no content, only five mentioned users. When viewed as graph, the tweet nodes have five \texttt{mention} edges but content value is \texttt{null}. This single tweet is interesting because contentless tweets are statistically rare. Now imagine that a larger tweet graph has small pockets of dense subgraphs consisting of contentless tweets. These subgraphs can be considered ``interesting'' both because of their lack of content and because of the high clusteredness. In contrast, the second tweet in Table \ref{tab:tweets} has content discussing the rapper ``TI" in a closed group. Even if there are similarly dense subgraphs representing an intense discussion on the theme, it is not necessarily interesting, unless the content of the conversation is very different from the content of the conversation of the graph surrounding it, which would be the case if everyone else in a network discusses politics while this group discusses a rapper or if the content of this group is extremely narrow. A version of this problem was investigated in \cite{zheng2019social}, focused on ``extreme tweeters'', users  who tweet a lot, have a very close but highly active network, but have a very limited discussion vocabulary. Thus, the notion of interestingness depends both on the content and structure of the subgraph and can only be interpreted in the context of a reference network as determined by an analyst's need.  

The goal of this paper is to discover all such interesting subgraphs from a social network graph where the context against which the interestingness search would be conducted is specified as queries against a graph database that materializes a social media network. We believe that discovering interesting subgraphs during the exploratory analysis of social media would reveal a pattern of user behavior that ``stands out'' and merits a more detailed follow-up investigation.
\begin{table*}[t]
\begin{tabular}{|l|l|l|}
\hline
  & \multicolumn{1}{|c|}{\textbf{Interesting Text from Tweets}}                                       & \multicolumn{1}{|c|}{\textbf{Why These Tweets are Interesting}}   \\ \hline
1 & @rooseveltinst @Justice4ADOS @SandyDarity @IrstenKMullen @MusicNegrito                                                                                                                                                                                                                                                                                                                                                                                                                                                         & \begin{tabular}[c]{@{}l@{}}Creating a strongly connected network \\ by mentioning only users like a robot.\end{tabular}
                                                                                                                           \\ \hline
2 & \begin{tabular}[c]{@{}l@{}}@noirdosser @chelleter\_d @SandyDarity @quantumblackne2 \\ @Tip @KeishaBottoms @esglaude I think \\ TI is fake shook..... typical move celebrities play.\end{tabular}                                                                                                                                                                                                                                                                                                                               & \begin{tabular}[c]{@{}l@{}}Creating a close issue centric network by \\ adding known and focused users. \end{tabular}
                                                                                                                     \\ \hline
3 & \begin{tabular}[c]{@{}l@{}}@princss6 @DerrickNAACP I agree. \\ At this critical juncture when the natl attention is on injustice to \\ \#ADOS he is “all black lives” mattering our justice claim. \\ This makes no sense. \#ResignDerrick\end{tabular}                                                                                                                                                                                                                                                                        &    \begin{tabular}[c]{@{}l@{}}While the content is simple, this tweet bridges two\\ different dense subnetworks by  co-mentioning two popular \\ users from these two networks. \end{tabular}                                                                                                                      \\ \hline
4 & \begin{tabular}[c]{@{}l@{}}@Hub\_Libertarian @davidenrich @realDonaldTrump \\ @DeutscheBank Love how ignored the facts about Supreme Court decisions...lol. 9-0, \\ the most common decision is facts you can’t ignore.\end{tabular}                                                                                                                                                                                                                                                                                           & \begin{tabular}[c]{@{}l@{}} Tweets like this are not interesting. They create a focused but\\ broad  network by mentioning all related users, some \\of whom are very popular. \end{tabular}                                                                                                           \\ \hline
5 & \begin{tabular}[c]{@{}l@{}}@grey\_geena @obiora\_odi @KHiveQueenBee @livemusic4me \\ @Cat\_MarqueeLV @Unknwnstuntman\\  @ElMcClelland @annableigh @thatboybesangin @fourgunfire @moshimisen\\  @sheanabana @twobesure @Alysson @NancyTabak @JoeBiden \\ We have no choice but to let it play out, however, \\ white folks out her writing letters to the manager \\ and equating life long Black public servants to “cosmetics” and “tokens”,\\  sooo yeah my trust in “the process” is minimal, right about now.\end{tabular} & \begin{tabular}[c]{@{}l@{}} Creating a broad network by mentioning \\ as many users as possible. \end{tabular}                                             \\ \hline
6 & \begin{tabular}[c]{@{}l@{}}@KBULTRA0  @KamalaHarris Tomorrow \\ I will conduct myself the way an old Italian \\ Catholic nona in Napoli celebrates Shivaratri\\ \textasciicircum this is the only "resistance" possible \\ In fact I`ve already partially ruined it\end{tabular}                                                                                                                                                                                                                                               & \begin{tabular}[c]{@{}l@{}}These types of tweet are interesting because they gain \\ attention by mentioning popular users who are \\ fairly unrelated to the content of the tweet \end{tabular} \\ \hline
\end{tabular}
\caption{Some types of tweets that are more ``interesting'' than others because the network around these tweets show some unusual phenomena (see text for more explanation).}
\label{tab:tweets}
\end{table*}

\section{Interesting Subgraphs of a Social Network}
\label{sec:formulation}
\noindent \textbf{Related Work.} The problem of finding interesting subgraphs has been investigated from several different viewpoints. One of the earliest ``graph mining'' approaches focused on discovering the most frequently occurring subgraphs \cite{lee2010survey,kuramochi2001frequent}. A second approach considers interesting subgraphs as a subgraph matching problem \cite{gupta2014top, he2017misaga, shan2019dynamic}. Their general approach is to compute all matching subgraphs that satisfy a user the query and then ranking the results based on the rarity and the likelihood of the associations among entities in the subgraphs. A third approach \cite{van2016subjective, adriaens2019subjectively} uses the notion of ``subjective interestingness'' which roughly corresponds to finding subgraphs whose connectivity properties (e.g., the average degree of a vertices) are distinctly different from an ``expected'' \textit{background} graph. Like many machine learning techniques, this approach uses a constrained optimization problem that maximizes an objective function over the \textit{information content} and the \textit{description length} of the desired subgraph pattern.

\noindent \textbf{Our Approach.} This work is inspired by the query-driven and subjective interestingness approaches. We assume that the social media is represented by an social media graph $G_0$. Like the query-driven approach, we initiate the discovery process by user-specified query $Q$ that identifies an initial subnetwork $G'=Q(G_0)$, called the \textit{initial background graph}, over which the discovery process is conducted. Further, like the ``subjective interestingness'' approach we discover subgraphs $S_i \subset G'$ whose content and structural features are distinctly different that of $G'$. However, unlike previous approaches, we apply a generate-and-test paradigm for discovery. The generate-step (Section \ref{sec:generate}) uses a graph cube like \cite{zhao2011graph} technique to generate candidate subgraphs that might be interesting and the test-step (Section \ref{sec:testing}) computes whether the candidate is sufficiently distinct from the initial background graph, and whether the collection of candidates are sufficiently distinct from each other.

\noindent \textbf{Subgraph Interestingness.} For a subgraph $S_i$ to be considered as a candidate, it must satisfy the following conditions.
\begin{enumerate}[label=(\roman*),align=left, leftmargin=*]
    \item \textbf{C1.} $S_i$ must be connected and should satisfy a size threshold $\theta_n$, the minimal number of nodes
    \item \textbf{C2.} Let $A_{ij}$ (resp. $B_{ik}$) be the set of \textit{local} properties of node $j$ (resp. edge $k$) of subgraph $S_i$. A property is called ``local'' if it is not a network property like vertex degree. All nodes (resp. edges) of $S_i$ must satisfy some user-specified predicate $\phi_N$ (resp. $\phi_E$) specified over $A_{ij}$ (resp. $B_{ik}$). 
    
    For example, a node predicate might require that all nodes of type ``post'' in the subgraph must have a re-post count of at least 300, an edge predicate may require that all hashtag cooccurrence relationships must have a weight of at least 10.
    
    The rationale for imposing a user defined constraint on the candidate subgraph is to improve the interpretability of the result. Typical subjective interestingness techniques \cite{van2016subjective, adriaens2019subjectively} use only structural features of the network and do not consider attribute-based constraints, which limits their pragmatic utility.
     
    \item \textbf{C3.} For each text-valued attribute $a$ of $A_{ij}$, let $C(a)$ be the collection of the values of $a$ over all nodes of $S_i$, and $\mathcal{D}(C(a))$ is a textual diversity metric computed over $C(a)$. For $S_i$ to be interesting, it must have at least one attribute $a$ such that $\mathcal{D}(C(a))$ does not have the usual power-law distribution expected in social networks. 
  
    Zheng et al \cite{zheng2019social} presents two such measures over tweet text -- vocabulary diversity (distribution of distinct non-stop-word terms) and topic diversity (computed as SVD vectors). They showed that interesting tweets show a significantly low diversity compared to those of ``standard'' tweet collections.
\end{enumerate}

\section{The Generate and Test Process}
\subsection{Candidate Generation}
\label{sec:generate}
\noindent \textbf{Initial Query.} The candidate generation process starts with an initial query $Q$ to the social network graph. The query is placed against the original social media data without considering their network structure. For example, a query can select all tweets containing the hashtag \texttt{\#ADOS}\footnote{American Descendant of Slaves} starting in 2019. The resulting collection becomes the universe of discourse for interestingness discovery. The initial background graph $G'$ is constructed on the results of this query.

\noindent \textbf{Node Grouping.} Given the graph $G'$, the user specifies a grouping condition expressed as a graph pattern. For example, the grouping pattern,  \texttt{(:tweet\{date\})-[:uses]->(:hashtag\{text\})}, expressed in a Cypher-like syntax \cite{francis2018cypher} (implemented in the Neo4J graph data management system) states that all tweets having the same posting date, together with every distinct hashtag text will be placed in a separate group. Notice that this process produces a ``soft'' partitioning on the tweets and hashtags due to the many-to-many relationship between tweets and hashtags. Hence, the same tweet node can belong to two different groups because it has multiple hashtags. Similarly, a hashtag node can belong to multiple groups because tweets from different dates may have used the same hashtag. While the grouping condition specification language can express more complex grouping conditions, in this paper we will use simpler cases to highlight the efficacy of the discovery algorithm. We denote the node set in each group as $N_i$.

\noindent \textbf{Graph Construction.} The graph construction phase constructs a subgraph $S_i$ by expanding on the node set $N_i$. Different expansion rules can be specified, leading to the formation of different graphs. Here we list three rules that we have found fairly useful in practice.
\begin{enumerate}[label=(\roman*), align=left, leftmargin=*]
    \item \textbf{G1.} Identify all the \texttt{tweet} nodes in $N_i$. Construct a \textit{relaxed induced subgraph} of the \texttt{tweet}-labeled nodes in $N_i$. The subgraph is induced because it only uses tweets contained within $N_i$, and it is \textit{relaxed} because contains all nodes \textit{directly associated} with these tweet nodes. These nodes include the tweet author, the hashtags and URLs contained in the tweets, the users mentioned in a tweet, Consequently the graph identifies the shared hashtags and user mentions.
    \item \textbf{G2.} Construct a \textit{mention network} from within the tweet nodes in $N_i$ -- the mention network initially connects all \texttt{tweet} and \texttt{user}-labeled nodes. Extend the mention network by including all nodes \textit{directly associated} with these tweet nodes. Notice that this \textit{relaxed induced subgraph} is a constrained version of the previous construction where only the \textit{mention} edge is considered.
    \item \textbf{G3.} A third construction relaxes the grouping constraint. We first compute either \textbf{G1} or \textbf{G2}, and then extend the graph by including the first order neighborhood of mentioned users or hashtags. While this clearly breaks the initial group boundaries, a network thus constructed includes tweets of similar themes (through hashtags) or audience (through mentions).
\end{enumerate}
Once these candidate graphs are constructed, they are tested for criterion \textbf{C3}. In this paper, we have directly applied the diversity metric proposed in \cite{zheng2019social}.
\subsection{Testing for Relative Interestingness}
\label{sec:testing}
In our setting, the interestingness of a subgraph is computed in reference to a background graph $G'$, and consists of a structural as well as a content component. We first discuss the structural component. To compare a subgraph $S_i$ with the background graph, we use $f(P_j(S_i)$, the frequency distribution $f(.)$ of the network properties $P_j$ of $S_i$ with that of the background and compute the difference of their distributions. The network properties we compute include different centrality measures while the distributions are compared based on their Jensen–Shannon divergence (JSD), which is a  symmetric and smoothed version of the Kullback–Leibler divergence measure to compare distributions. In the following, we use $\Delta(a,b)$ to refer to the JS-divergence of two distributions $a$ and $b$. 

\begin{itemize}[leftmargin=*]
    \item \textbf{High-Centrality Nodes:} The testing process starts by identifying the distributions of nodes with high node centrality between the networks. While there is no shortage of centrality measures in the literature, we choose eigenvector centrality, defined below, to represent the dominant nodes. 
    Let $A = (a_{i,j})$ be the adjacency matrix of a graph. The eigenvector centrality $x_{i}$ of node $i$ is given by: $$x_i = \frac{1}{\lambda} \sum_k a_{k,i} \, x_k$$ where $\lambda \neq 0$ is a constant. 
    The rationale for this choice follows from earlier studies in \cite{Bonacich2007-mx,Ruhnau2000-jy,Yan2014-dn}, who establish that since the eigenvector centrality can be seen as a weighted sum of direct and indirect connections, it represents the true structure of the network more faithfully than other centrality measures. Further, \cite{Ruhnau2000-jy} proved that the eigenvector-centrality under the Euclidean norm can be transformed into node-centrality, a property not exhibited by other common measures.

    Let the distributions of eigenvector centrality of subgraphs $A$ and $B$ be $\beta_a$ and $\beta_b$ respectively, and the distribution of the background graph is $\beta_t$, then $$\Delta_e(\beta_t, \beta_a)>\Delta_e(\beta_t, \beta_b)$$ indicates that subgraph $A$ contains more influnetial nodes then subgarph $B$. 

    \item \textbf{Navigability:} The second network feature we consider is \textit{edge betweenness centrality} defined below.   Let $\alpha_{ij}$ be the number of shortest paths from node i to j and $\alpha_{ij}(k)$ is the number of paths passes through the edge $k$. Then the edge-betweenness centrality is $$C_{eb}(k)= \sum_{(i,j)\in V} \frac{\alpha_{ij}(k)}{\alpha_{ij}}$$
    By this definition, the edge betweenness centrality is the portion of all-pairs shortest paths that pass through an edge. Our choice of edge betweenness centrality stems from the observation that and subgraph with a higher proportion of high-valued edge betweenness centrality implies that a this subgraph is more \textit{navigable} than the rest of the graph, i.e., information propagation is higher through this subgraph compared to the whole background network, for that matter, any other subgraph of network having a lower proportion of nodes with high edge betweenness centrality. Let the distribution of the edge betweenness centrality of two subgraphs $A$ and $B$ are $c_1$ and $c_2$ respectively, and the edge betweenness centrality distribution of the reference graph is $d$. Then, $$\Delta_b(d, c_1) < \Delta_b(d, c_2)$$  means the second subgraph is more navigable than the first subgraph.
    
    \item \textbf{Propagativeness:} The navigability of a network determines the coverage of the information flow, but does not determine the propagation movement within the network. We use current flow betweenness centrality and the average neighbor degree jointly to determine the possibility of the higher rate of propagation within the network. The current flow betweenness centrality is the portion of all-pairs shortest paths that pass through a node, and the average neighbor degree is the average degree of the neighborhood of each node. If a subgraph has higher current flow betweenness centrality plus a higher average neighbor degree, the network should have faster communicability.
    \\ 
    Let $\alpha_{ij}$ be the number of shortest paths from node $i$ to $j$ and $\alpha_{ij}(n)$ is the number of paths passes through the node $n$. Then the current flow betweenness centrality: $$C_{nb}(n)= \sum_{(i,j)\in V} \frac{\alpha_{ij}(n)}{\alpha_{ij}}$$
    \\
    Suppose the distribution of the node betweenness centrality of two subgraphs $A$ and $B$ is $p_1$ and $p_2$ respectively, and distribution of the reference graph is $p_t$. Also the distribution of the $\beta_{n}$, the average neighbor degree of the node $n$, for the subgraph $A$ and $B$ is $\gamma_1$ and $\gamma_1$ respectively, and the true distribution is $\gamma_t$. If the condition
    $$\Delta(p_t, p_1) + \Delta(\gamma_t, \gamma_1) < \Delta(p_t, p_2) +  \Delta(\gamma_t, \gamma_2)$$
    holds, we can conclude that subgraph $B$ can be deemed as a faster propagating network than subgraph $A$. This measure is of interest in a social media based on the observation that misinformation/disinformation propagation groups either try to increase the average neighbor degree by adding fake nodes or try to involve influential nodes with high edge centrality to propagate the message faster \cite{besel2018full}.  
    
    \item \textbf{Subgroups within a Candidate Subgraph:} The last metric relates to the diversity of groups within a candidate interesting subgraph based on the above criteria. The number of subclusters within a candidate subgraph depicts whether the subgraph should be further decomposed into smaller subgraphs that would signify a finer-grain identification of interest zones. To this end, we use subgraph centrality and coreness of nodes as our metrics. The subgraph centrality $SC(i)$ of a vertex $i$ is given by
    $$ SC(i) = \sum_{k=0}^\infty \frac{\mu_k(i)}{k!}$$
    where $\mu_k(i)$ is a local spectral moments defined as the $i$-th diagonal entry of the $k$-th power of the graph's adjacency matrix \cite{estrada2005subgraph}. The subgraph centrality measures the number of subgraphs a vertex participates and the core number of a node is the largest value $k$ of a $k$-core containing that node. So a subgraph for which the core number and subgraph centrality distributions are right-skewed compared to the background subgraph are (i) either split around high-coreness nodes, or b) reported to the user as a mixture of  diverse topics. 
\end{itemize}

\section{The Discovery Process}
\label{sec:discovery}
Based on the metrics and the general principles presented in the previous subsections, the interesting subgraph discovery process is implemented through two algorithms. Algorithm \ref{alg:graph-metrics} constructs the graph, and while Algorithm \ref{alg:discovery-algo} discovers the exciting patterns progressively. 

\begin{algorithm}
\caption{Graph Construction Algorithm}
\label{alg:graph-metrics}
\SetKwProg{ComputeMetrics}{Function \emph{ComputeMetrics}}{}{end}
INPUT : $Q_{out}$ Output of the query, $L$ Graph construction rules, $gv$ grouping variable, $th_{size}$ is the minimum size of the subgraph\;
\SetKwProg{gmetrics}{Function \emph{gmetrics}}{}{end}
\SetKwProg{CompareHistograms}{Function \emph{CompareHistograms}}{}{end}
\gmetrics{($Q_{out}$, $L$, $groupVar$)}{
G[]$\leftarrow$ ConstructGraph($Q_{out}$, $L$)\;
$T \leftarrow$ []\;
\For{$g \in G $}{
     $t_{\alpha} \leftarrow$ ComputeMetrics(g)\; 
     $T.push(t_{alpha})$\;
    
    }
return $T$
}
\ComputeMetrics{(Graph g)}{
$m\leftarrow[]$\;
$m.push(eigenVectorCentrality(g))$\;
.........
$m.push(coreNumber(g))$\;
return $m$
}
\CompareHistograms{(List $t_{1}$, List $x_{2}$)}{
$s_g \leftarrow cut2bin(x_2, bin_{edges})$\;
$bin_{edges} \leftarrow$ getBinEdges($x_{2}$)\;
$t_g \leftarrow cut2bin(t_1, bin_{edges})$\;

$\beta_{js} \leftarrow distance.jensenShannon(t_g, s_g)$\;
$h_t \leftarrow histogram(t_g, s_g,bin_{edges} )$\;
return $\beta_{js}, h_t, bin_{edges}$\;
}
\end{algorithm}

\noindent \textbf{Graph Construction Algorithm: }
Recall that the query output $Q_{out}$ is the result of the query performed against the social media data with a set of filtering keywords, and without considering the network topology.

\noindent Graph construction starts with the $gmetrics$ function of the algorithm \ref{alg:graph-metrics}. The inputs of the \texttt{gmetrics} algorithm are (a) the output of the user's initial query, (b) graph construction rules (e.g., induced subgraph), and (c) the grouping variable(s). Although one can use multiple grouping variables for the algorithms, the following presentation assumes, with no loss of generality, a single grouping variable.

\noindent The construction rules used by the algorithm are specified through views (i.e., rules) that construct edges by evaluating path expressions.  For example, consider the rule: 
\begin{multline*}
(a:user)-[:mentions]\rightarrow(b:user) ~~~\textbf{if} \\   
                  (a)-[authors]\rightarrow(t:tweet)-[:mentions]\rightarrow(b:user). 
\end{multline*}
The LHS of the rule is the edge constructed in the result graph if the RHS is satisfied. This rule constructs an edge labeled \texttt{mentions} from node $a$ to node $b$, both of the type user such that user $a$ has authored a tweet that mentions a user $b$ in the same tweet. Furthermore, the rule set controls the different construction phases explained in the Section \ref{sec:generate}. Depending on the rules, it can construct an induced subgraph or a relaxed subgraph. 

\noindent The algorithm uses the grouping variable $gv$ to create a soft partitioning over the over the set of vertices and apply the graph construct rules to construct the graph.
After the graph formation, the threshold value is used to filter out the smaller subgraphs before passing it to the $ComputerMetrics$ function. The $ComputerMetrics$ function takes each subgraph as an input, and computes a set of centrality measurements on the subgraph. The $ComputerMetrics$ returns a list of centrality values for each node for the subgraph. We are currently computing six centrality measures, viz. Eigenvector Centrality, edge current flow betweenness centrality, subgraph centrality, and current flow betweenness centrality. Additionally, we are also calculating the average neighbor degree and the core number for each node. 

\noindent The output of each metric produces a value for each participant node of  the input. However, to compare two different candidates, in terms of the metrics mentioned above,  we need to convert them to comparable histograms by applying a binning function depending on the data type of the grouping function. 

\noindent \textit{Bin Formation (cut2bin):} Cut is a conventional operator (available with R, Matlab, Pandas etc. ) segments and sorts data values into bins. The cut2bin is an extension of a standard cut function, which compares the histograms of the two distributions who domains (X-values) must overlap. The cut function accepts as input a set of set of node property values (e.g., the centrality metrics), and optionally a set of edge boundaries for the bins. It returns the histograms of distribution. Using the cut, first, we produce $n$ equi-width bins from the distribution with the narrower domain. Then we extract bin edges from the result and use it as the input bin edges to create the wider distribution`s cut.  This enforces the histograms to be compatible. In case one of the distribution is known to be a reference distribution (distribution from the background graph) against which the second distribution is compared, we use the reference distribution for equi-width binning and bin the second distribution relative to the first.\\
\noindent The $CompareHistograms$ function uses the \textit{cut2bin} function to produce the histograms, and then computes the JS Divergence on the comparable histograms. The $CompareHistograms$ function returns the set of divergence values for each metric of a subgraph, which is the input of the discovery algorithm. The function requires the user to specify which of the compared graphs should be considered as a reference -- this is required to ensure that our method is scalable for large background graphs (which are typically much larger than the interesting subgraphs). If the background graph is very large, we take several random subgraphs from this graph to ensure they are representative before the actual comparisons are conducted. To this end, we adopt the well-known random walk strategy.
In the experiments, we used three random walks to introduce sufficient randomness. \\
Hence, the output of the $CompareHistograms$ is JS-divergence value for each candidate respect to these random samples of a common reference graph. \\  
\begin{algorithm}
\caption{Graph Discovery Algorithm}
\label{alg:discovery-algo}
\SetKwProg{discover}{Function \emph{discover}}{}{end}
\KwIn{ Set of all subgraphs divergence $\sigma$}
\KwOut{Feature vectors $v_1$, $v_2$, $v_3$, List for re-partition recommendations $l$}
$ev$ : eigenvector centrality\;
$ec$ : edge current flow betweenness centrality\;
$nc$ : current flow betweenness centrality\;
$\mu$ : core number\;
$z$ : average neighbor degree\;
\discover{($\sigma$)}{
\For{any two set of divergence from $\sigma_1$ ans $\sigma_2$}{
\If{$\sigma_2(ev) > \sigma_1(ev)$}{
 $v_1(\sigma_2) = v_1(\sigma_2) + 1$\;
 \If{$\sigma_2(ec) > \sigma_1(ec)$}{
 $v_2(\sigma_2) = v_2(\sigma_2) + 1$\;
 \If{($\sigma_2(nc)+ \sigma_2(\mu)) > (\sigma_1(ec) + \sigma_2(\mu)$)}{
 $v_3(\sigma_2) = v_3(\sigma_2) + 1$\;
  }
  \If{($\sigma_2(sc)+ \sigma_2(z)) > (\sigma_1(sc) + \sigma_2(z)$)}{
 $l(\sigma_2) = 1$\;
  }
 }
 }
 }
}
\end{algorithm} 
\noindent \textbf{Discovery Algorithm : }
The discovery algorithm’s input is the list of divergence values of two candidate sets computed against the same reference graph. It produces four lists at the end. Each of the first three lists contains one specific factor of interestingness of the subgraph. The most interesting subgraph should present in all three vectors. If the subgraph has many cores and is sufficiently dense, then the system considers the subgraph to be \textit{uninterpretable} and sends it for re-partitioning. Therefore, the fourth list contains the subgraph that should partition again. Currently, our repartitioning strategy is to take subsets of the original keyword list provided by the user at the beginning of the discovery process to re-initiate the discovery process for the dense, uninterpretable subgraph.\\
In the algorithm $v_1$, $v_2$ and $v_3$ are the three vectors to store the interestingness factors of the subgraphs, and $l$ is the list for repartitioning.  For two subgraphs, if one of them qualified for $v_1$ means, the subgraph contains high centrality than the other. In that case, it increase the value of that qualified bit in the vector by one. Similarly, it increases the value of  $v_2$ by one,  if the same candidate has high navigability. Finally, it increases the $v_3$, if it has higher propagativeness.  After the execution of all combinations of the candidate,  it selects the top-k scored of candidates from each vector and marks them interesting.  

\section{Experiments and Results}
\label{sec:experiment}

\noindent \textsc{Dataset:} The experimental dataset was gathered in the following manner. 1) We collected a set of tweets over a period of six months, such that the tweets use the the hashtag \texttt{\#ADOS}, usually associated with Black American issues; 2) We adopt a snowball sampling strategy by which we identify the most very active users based on the number of tweets they author; 3) We collect all tweets from these users regardless of the topic content;  4) This process is performed for two more rounds. From this set, we eliminate that uses only non-text symbols like emojis. The size of the accumulated dataset is 9,780,590 tweets, and the number of unique users mentioned is 89,8850.

\noindent As mentioned in Section \ref{sec:generate}, the candidate generation process starts with a keyword query to the tweet text, and we generated three different candidates from a different set of keywords. A list of the keywords and the name of the collection is given in in Table \ref{tab:group-tab}. The first column of the table is the group’s name, and the second column represents the group’s descriptions. In the candidate formation query, each group is represented by a set of keywords selected based on Google Trends such that these keywords cooccur with our seed keyword \texttt{ADOS}. 

\noindent \textit{Node Grouping:} Initially, for each candidate, we grouped them using the popularity count of the tweet and the followers’ count of the user. The grouping operation is implemented using a binning strategy called the ``cut'', discussed in Section \ref{sec:discovery}. For the purposes of this experiment, we have explored 10 different node groups, and the graph graphs are checked against our interestingness criteria.  Furthermore, we empirically determine that attributes the tweet-popularity is a suitable the soft grouping variable is significant and practical to analyze because the followers count does not relate the content or the event directly. Hence we continue the experiment with tweet’s popularity number as the grouping variable. 
\begin{table}[t]
\centering
\begin{tabular}{|l|l|l|} \hline
  & Group & Description                        \\ \hline \hline
1 & A     & \#ADOS Movement Related Group        \\ \hline
2 & B     & American Political Group           \\ \hline
3 & C     & General Black Related Issue        \\ \hline 
4 & D     & HIV, Drug etc. related             \\ \hline
5 & E     & LGBT and Gay Issues                \\ \hline
6 & F     & Random terms from Google top trends \\ \hline
\end{tabular}
\caption{List of Candidates with domain descriptions used in the Experiments.}
\label{tab:group-tab}
\end{table}

\begin{figure*}[t]

\begin{subfigure}{.33\textwidth}
  \centering
  \includegraphics[width=.9\linewidth]{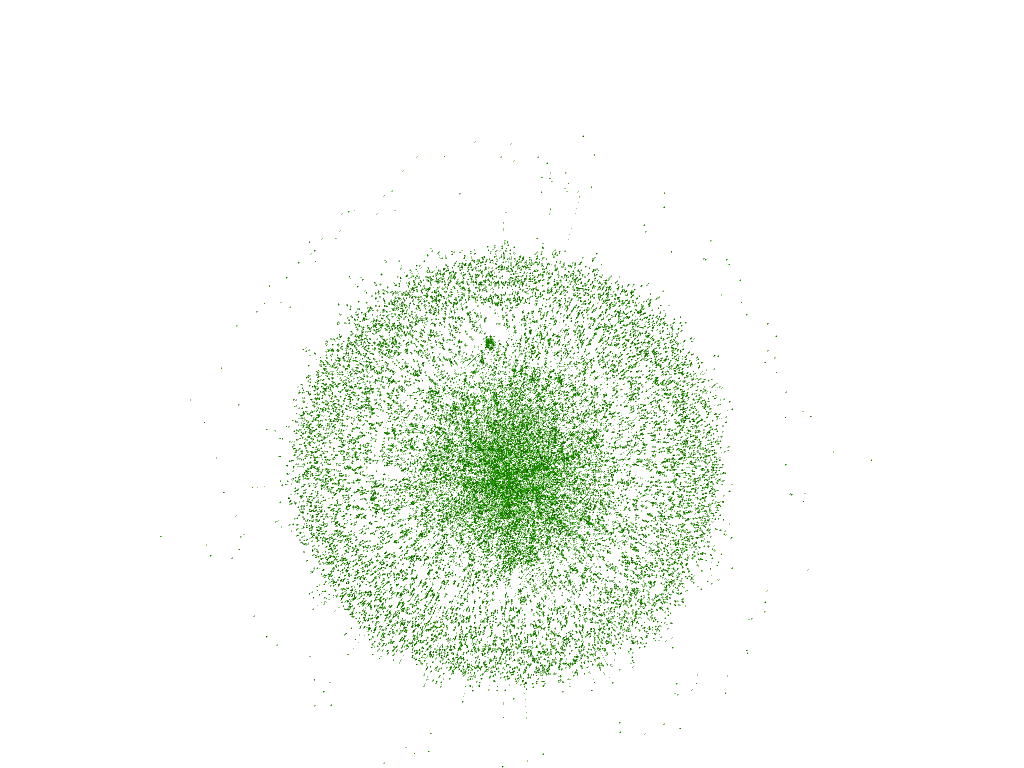}
  \caption{Random Sample of the graph -- 1}
 \label{fig:rwalk-1}
\end{subfigure}
\begin{subfigure}{.33\textwidth}
  \centering

  \includegraphics[width=.9\linewidth]{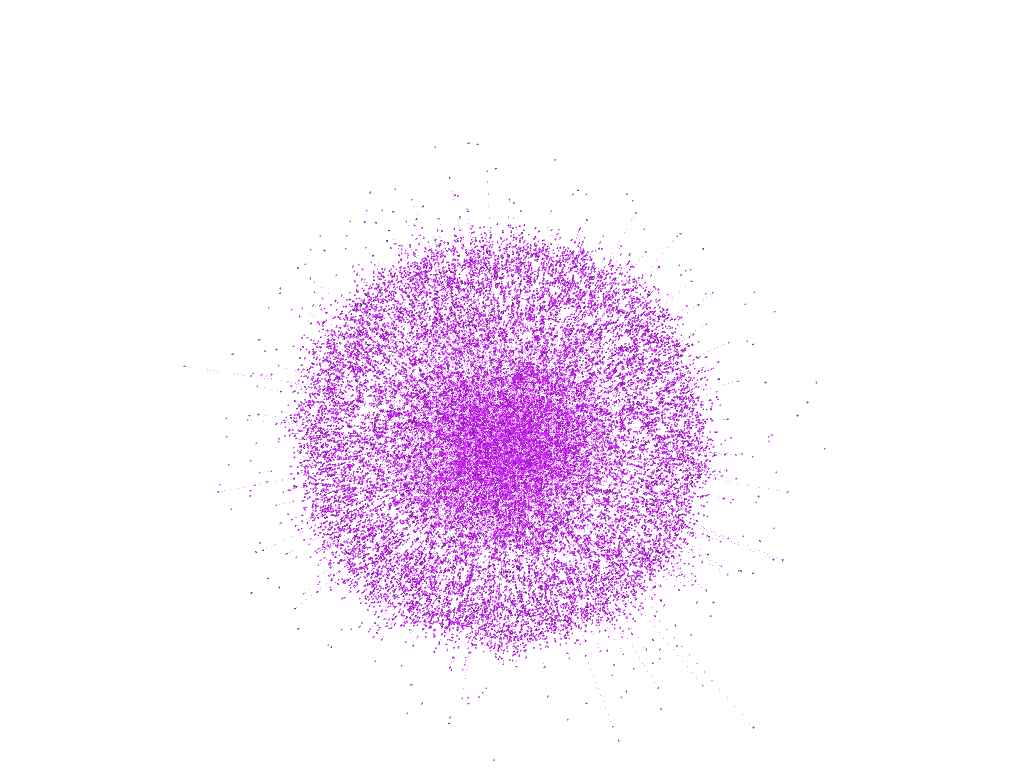}
  \caption{Random Sample of the graph -- 2}
 \label{fig:rwalk-2}
\end{subfigure}
\begin{subfigure}{.33\textwidth}
  \centering
  
  \includegraphics[width=.9\linewidth]{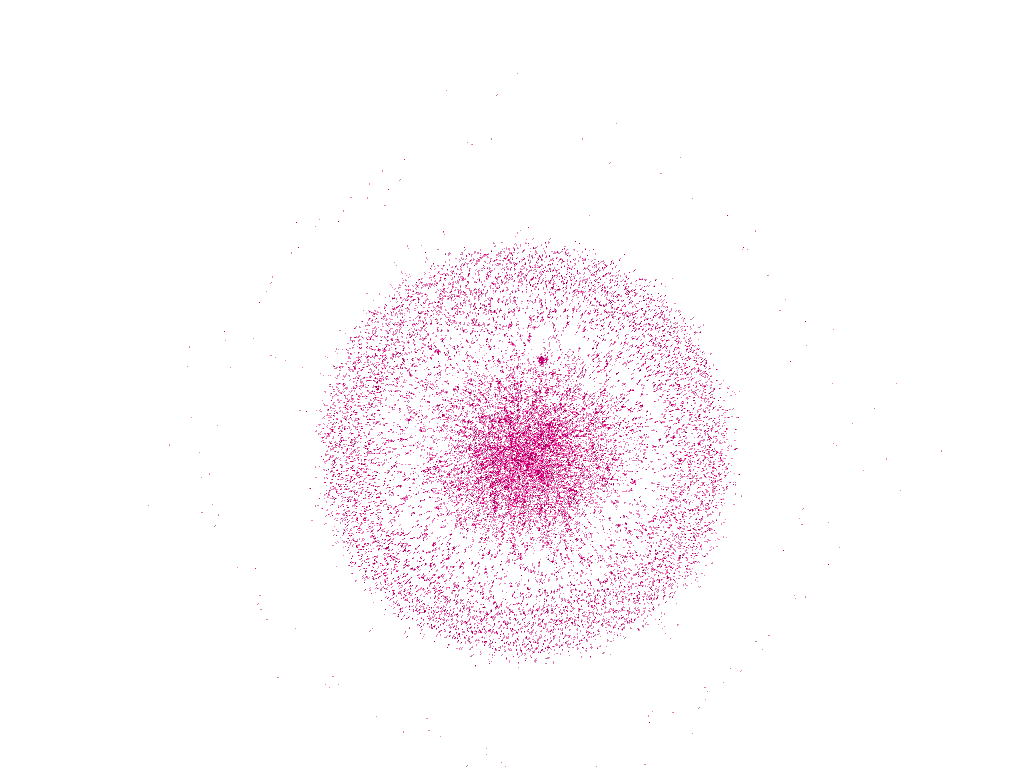}
  \caption{Random Sample of the graph -- 3.}
 \label{fig:rwalk-3}
\end{subfigure}

\begin{subfigure}{.3\textwidth}
  \centering
  \includegraphics[width=.9\linewidth]{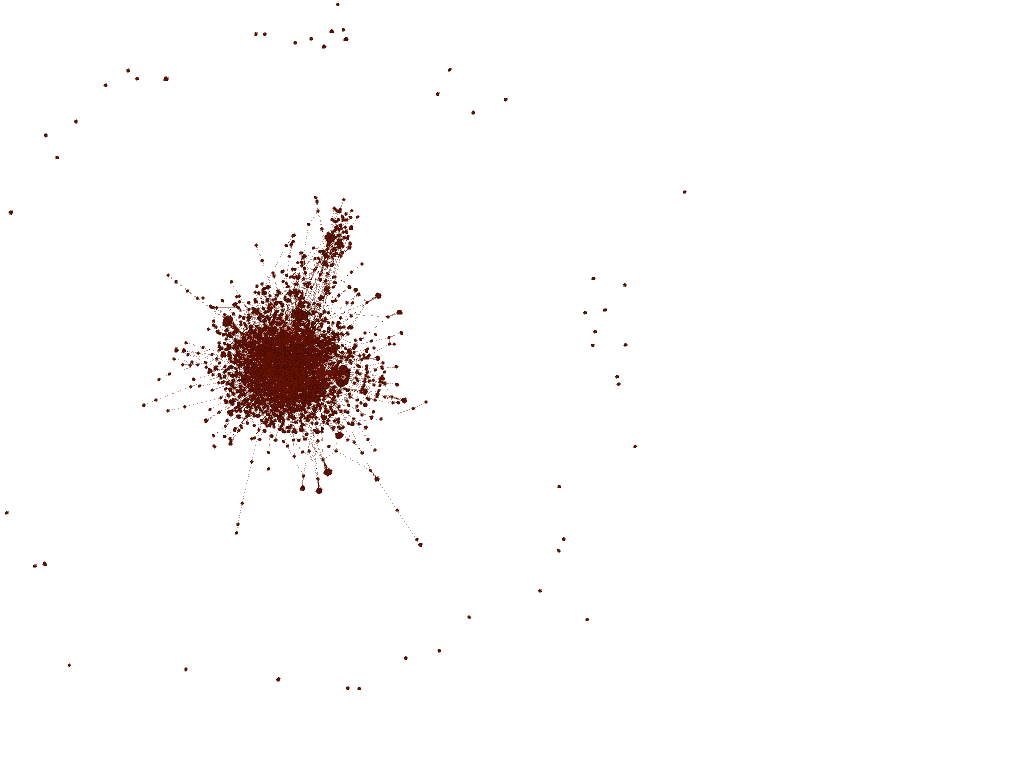}
  \caption{\#ADOS Movement Related Network filtered using \#ADOS related Keywords. 
}
  \label{fig:ados}
\end{subfigure}%
\begin{subfigure}{.33\textwidth}
  \centering
  \includegraphics[width=.9\linewidth]{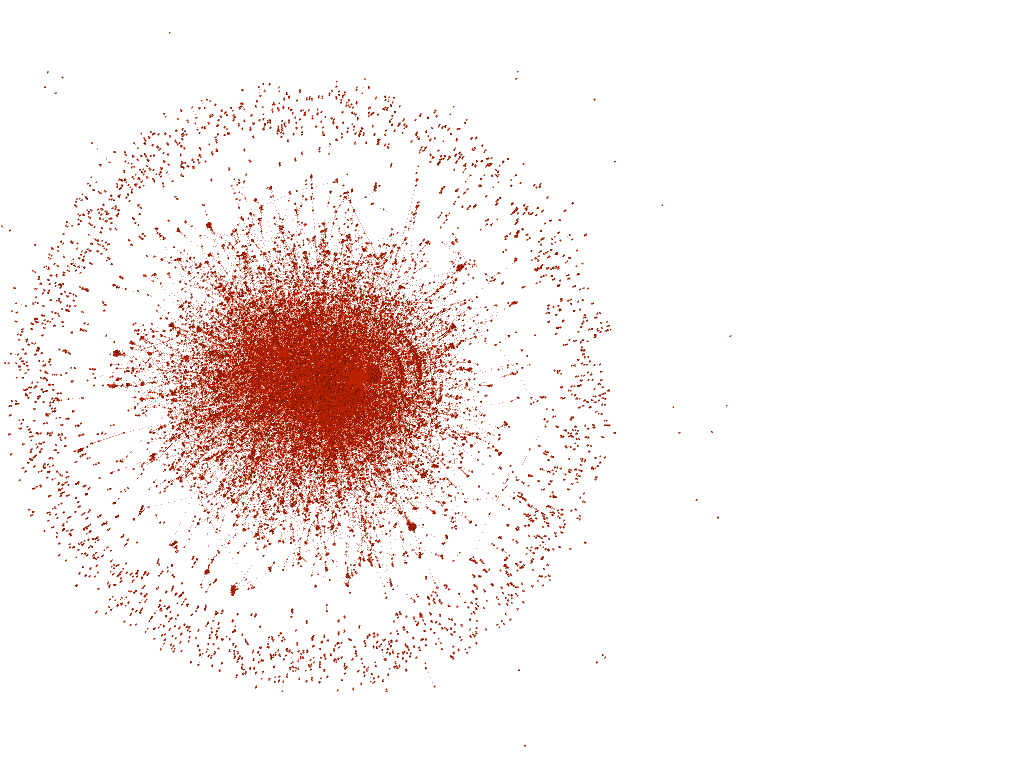}
  \caption{Political campaign-related network Based on the Presence of Political Personality.}
  \label{fig:political}
\end{subfigure}
\begin{subfigure}{.33\textwidth}
  \centering
  \includegraphics[width=.9\linewidth]{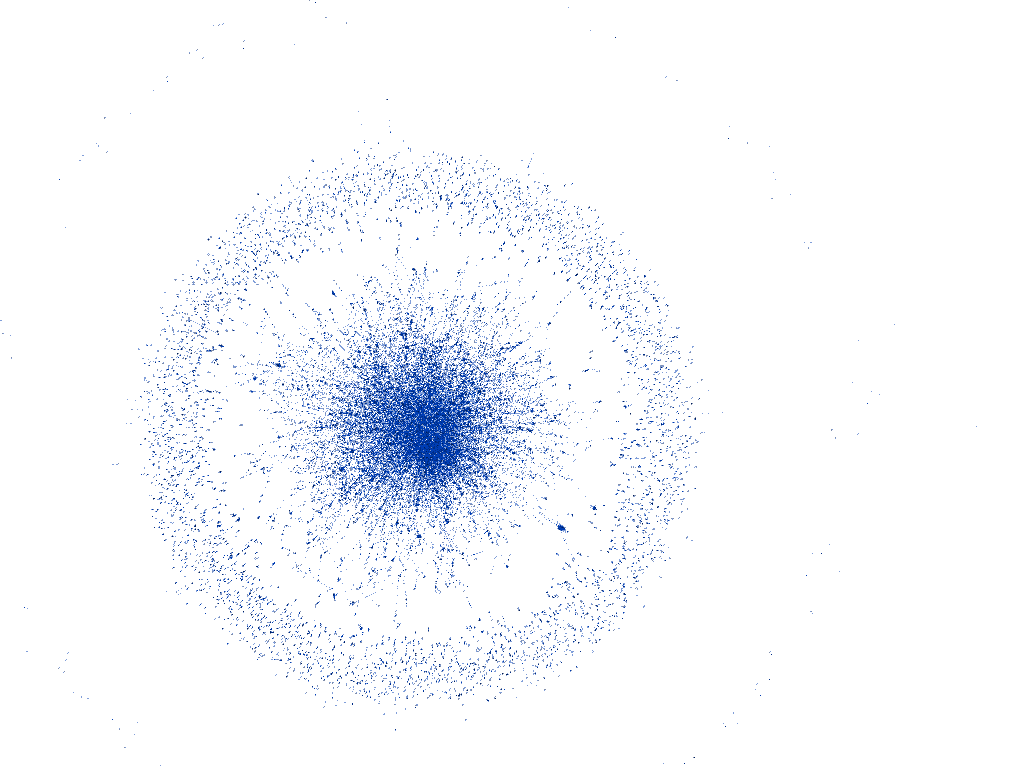}
  \caption{Black Social Issues Network.}
  \label{fig:gen-balck}
\end{subfigure}
\begin{subfigure}{.33\textwidth}
 
  \centering
  \includegraphics[width=1.1\linewidth]{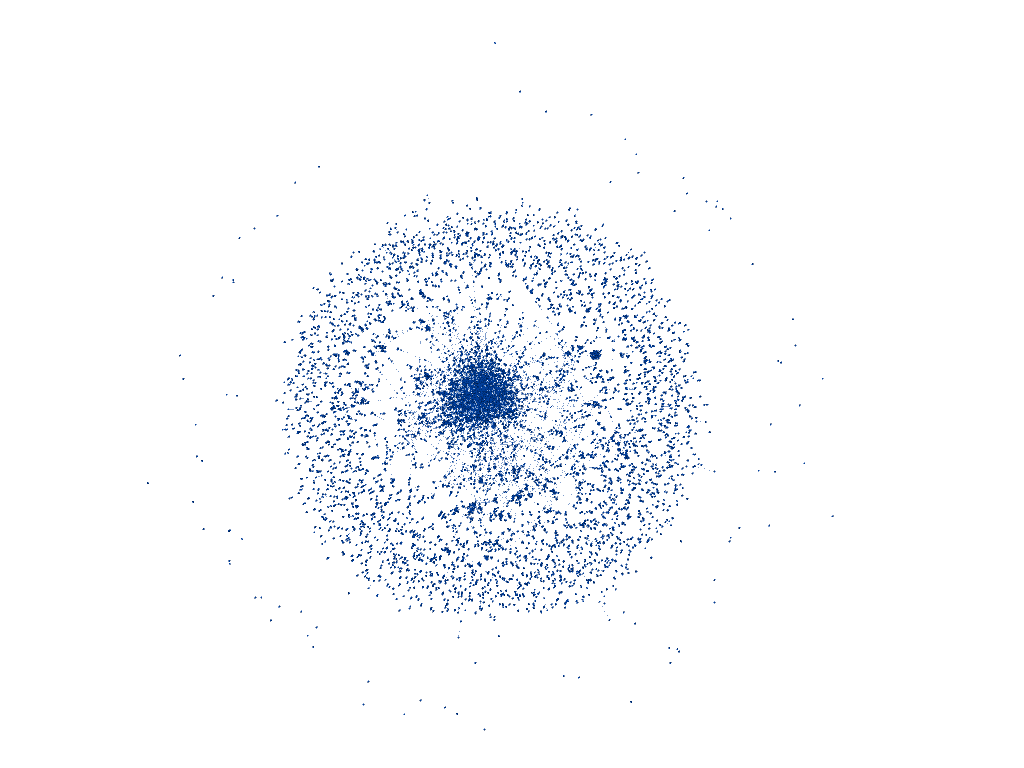}
  \caption{HIV, Drug and PrEP relates Issues.}
\label{fig:hiv}
\end{subfigure}
\begin{subfigure}{.33\textwidth}
  \centering
  \includegraphics[width=1.1\linewidth]{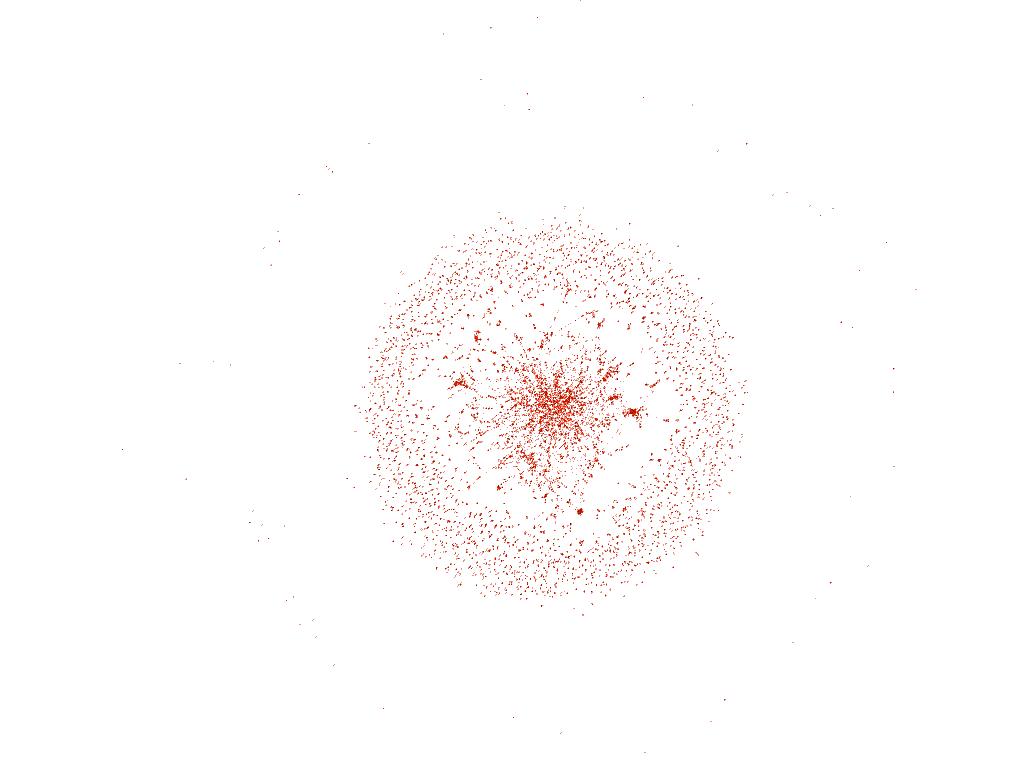}
  \caption{LGBTQ Community Related Group.}
  \label{fig:lgbt}
\end{subfigure}
\begin{subfigure}{.33\textwidth}
  \centering
  \includegraphics[width=1.1\linewidth]{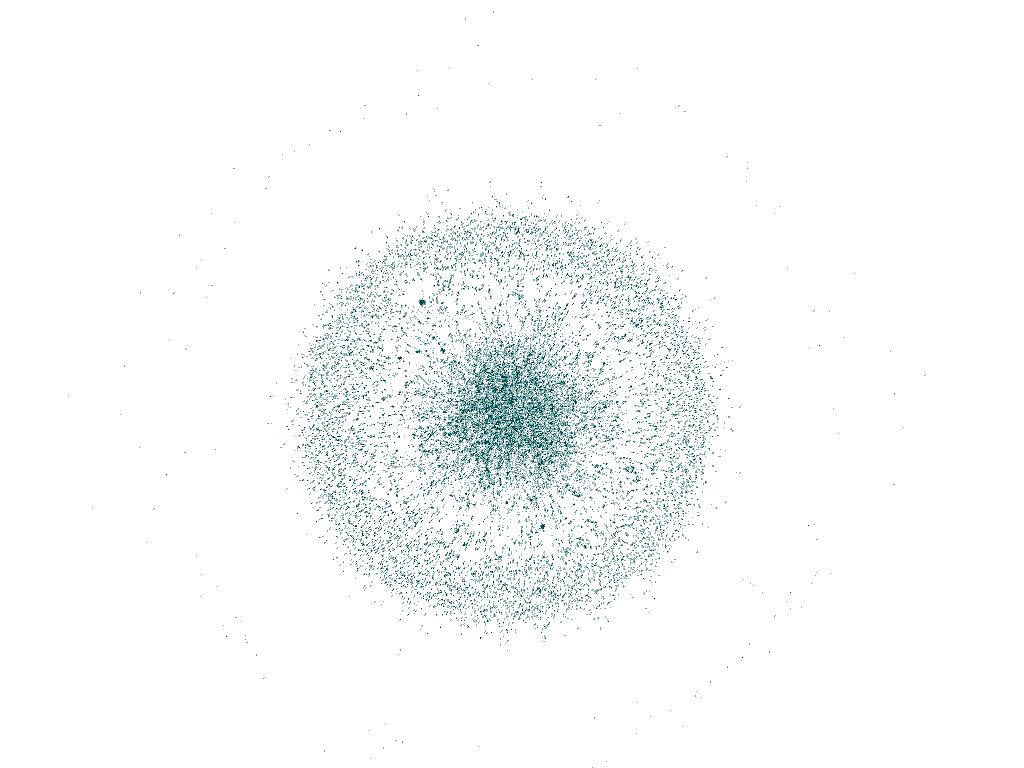}
  \caption{Random Terms from Google Trend}
  \label{fig:rand}
\end{subfigure}

\caption{User-Mention Network of Three different Data set. }
\label{fig:fig-net}
\end{figure*}

\begin{figure*}[t]
\begin{subfigure}{.35\textwidth}
   \centering
  \includegraphics[width=5.5cm , height=5cm]{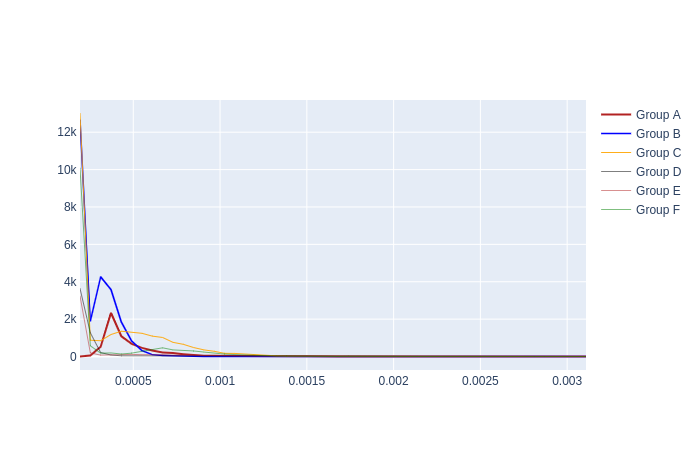}
  \caption{Distributions of navigability.}
  \label{fig:navi}
\end{subfigure}
\begin{subfigure}{.35\textwidth}
  \centering
  \includegraphics[width=5.5cm , height=5cm]{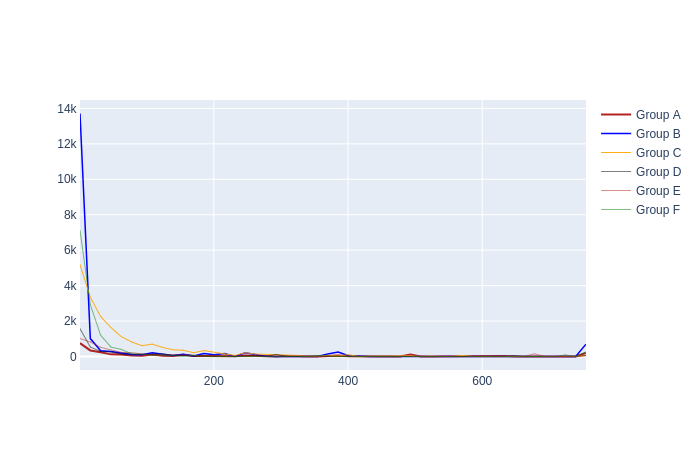}
  \caption{Distributions of propagativeness.}
  \label{fig:prop}
\end{subfigure}
\begin{subfigure}{.3\textwidth}
  \centering
  \includegraphics[width=5.5cm , height=5cm]{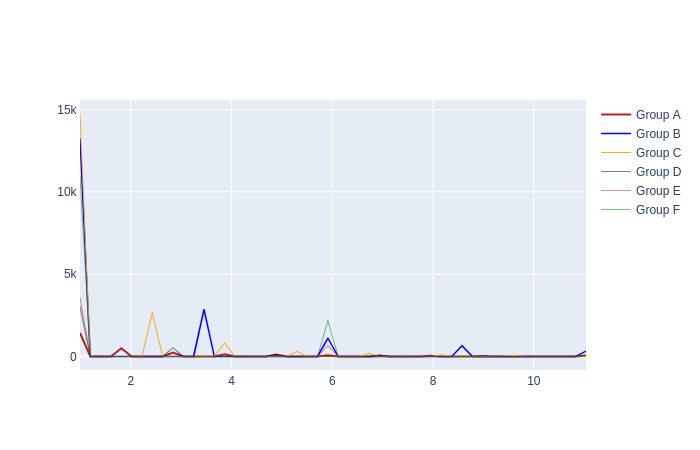}
  \caption{Distributions of Subgroups in Candidates.}
  \label{fig:subg}
\end{subfigure}
\caption{Comparative Distribution of all Candidates.}
\label{fig:all-garph}
\end{figure*}

\noindent \textsc{Experiments:}
We conducted experiments on the keyword categories shown in Table \ref{tab:group-tab}. In these categories, the first is directly related to the keywords used for data collection and the last one is randomly picked from Google trends with no relationship with the first. The remaining four have been selected as increasingly general issues found in Google Trends. Figure \ref{fig:fig-net} shows the network representation of the largest subgraph of each community, ad  Figures \ref{fig:rwalk-1}, \ref{fig:rwalk-2}, and \ref{fig:rwalk-3} shows three completely random fragments of our social network graph. Figures  \ref{fig:navi}, \ref{fig:prop}, and \ref{fig:subg} represent the outcome of the navigability, propagativenss and subgroup properties respectively. 

\noindent \textsc{Results:}
To present our interestingness results, we first present three random subnetworks sampled from our background graph. Recall, that while our seed for data collection indeed started with \texttt{\#ADOS}, we collected all tweets from users and their mention-neighbors recursively to reduce selection bias. The primary observation about these graphs is that although they do have a perceptible nucleus-periphery structure, the width of the ``peripheral ring'' is thick and the space between the nucleus and the periphery is fairly crowded. The sample network in Figure \ref{fig:rwalk-2} illustrates that some parts of the graph almost does not show any distinct peripheral boundary that establishes the strong edge formation probability between nucleus and non-nucleus nodes as well between a random pair of non-nucleus nodes.

\noindent Given this backdrop, let us examine the subgraphs shown in Figure \ref{fig:fig-net} -- they are examples of positive and negative results from our algorithm.

 \noindent \textbf{Subgraph 1.} The subgraph shown in Figure \ref{fig:ados}, characterized by a tight, strong core and a very scant periphery, is structurally interesting because it is significantly isolated from the rest of the network. Upon content analysis, it turns out to be strongly focused on ADOS issues with extremely high interactions amongst users who have very little interest outside this narrow scope. These users almost always mention only each other, participate in a meaningful conversation, and repeatedly use a restricted set of vocabularies and hashtags. The third tweet from Table \ref{tab:tweets} is the example of such tweets. In order to build a strong network community, they mention a small set of users numerous times, even without any content (first tweet of the same table). The signature of such a network is an intense core and very few nodes outside the nucleus. The eigenvector centrality distribution of such a network is higher than the random graph. The navigability will be relatively high (Group A in Figure \ref{fig:navi}), but due to less number of participant network will have average or low propagativeness (Group A in Figure \ref{fig:prop}). However, the likelihood of further subgroups within a subgroup is extremely low. \textit{We therefore conclude that Subgraph 1 is interesting.}\\
\noindent \textbf{Subgraph 2.} Figure \ref{fig:political}  or the political network is an example of an extensive network with a large and dense nucleus and a less dense but thicker periphery, which is not very strongly connected to the nucleus. Like tweet 4 from Table \ref{tab:tweets},  people in the center wish to connect to strongly connected and focused network by mentioning other connected users and issues. In tweet 5, people mention random unrelated users purposefully because it boosts their tweets' reach with loosely connected users, which creates a thick ring outside the kernel. We can also recognize such a network from very high navigability and propagativeness with comparatively fewer cores. From figures \ref{fig:navi}, \ref{fig:prop}, and \ref{fig:subg}, we can see that it has very high navigability and propagativeness compared to the other groups. Hence \textit{Subgraph 2 is interesting because it characterizes users who attempt to build bridges to promote message propagation}. \\
\noindent \textbf{Subgraph 3.} Figure \ref{fig:gen-balck} shows a network related to black issues (like healthcare) without specific focus on political issues. Hence, the network is not very intense (has a lighter nucleus), a peripheral density like Subgraph 2, and a diffuse space between then. Curiously, all our interestingness metrics score this subgraph highly. Upon closer inspection from  Figures \ref{fig:navi}, \ref{fig:prop}, and \ref{fig:subg}, we can see it has a spike on navigability, is well connected, and has a high propagativeness. The network also exhibits a high number of cores and subgroups \textit{Hence we label this subgraph as interesting but not readlily interpretable}. So this network is considered for further partitioning. \\
\noindent \textbf{Subgraphs 4 and 5.}. The networks shown in \ref{fig:hiv} and \ref{fig:lgbt} are based on a deliberate choice of ``general purpose'' topics. Clearly they have a lighter nucleus with a diffused ring, and fairly close to the random networks shown in the top 3 figures. This is confirmed by the low JS-divergence values for the navigability, propagativeness, and subgroup measures. \textit{Hence we conclude these three candidates are \textbf{not interesting} subgraphs} in our context.  \\
\noindent \textbf{Subgraph 6.} Finally a sample subgraph shown in Figure \ref{fig:rand}  produced from random set of keywords shows inconsistent results from our algorithm because the measures show no conclusive score on any one of the metrics that make it a proper interestingness candidate. In fact all subgraphs constructed from from our grouping operations produce any definitively interesting results. \textit{We therefore conclude that the groups from this set of keywords no significant difference with the background graphs, and are \textbf{not interesting}}.


\section{Conclusion}
\label{sec:conclusion}
This paper presents a novel algorithm in finding interesting subgraphs from a social network based on user's interests. We have combined the notions of graph grouping and subjective interestingness to create interestingness metrics and have evaluated them on a real-world data set. Our experiments show that the subgraphs that our algorithms report are indeed interesting. Our future work would involve making the algorithms more robust and devise a more elaborate evaluation methodology to validate the interestingness of the subgraphs recognized by our technique. We also intend to explore efficiency and scalability issues of the algorithm in future publications.

\noindent \textbf{Acknowledgment.} This work has been partially funded by NSF grants 1909875 and 1738411.

\bibliographystyle{IEEEtran}
\bibliography{paper.bib}
\end{document}